\documentclass[11pt]{article}
\usepackage{sigsam, amsmath}
\usepackage{multirow}
\usepackage{float}
\usepackage{graphicx}
\usepackage[colorlinks,linkcolor=blue,urlcolor=blue,citecolor=blue]{hyperref}
\usepackage{listings}

\DeclareMathOperator{\conv}{conv}
\newcommand{\spt}[1]{\text{spt}(#1)}

\issue{Vol.xx, No.xx, Issue xxx, month 201x}
\articlehead{ISSAC 2018 abstracts}
\titlehead{\textsc{RealCertify}: a Maple package for certifying non-negativity}
\authorhead{Victor Magron and Mohab Safey El Din}
\begin{document}

\title{\textsc{RealCertify}: a Maple package for certifying non-negativity}

\author{
Victor Magron\\
CNRS Verimag, Sorbonne Universit\'e, \textsc{INRIA},\\
Laboratoire d'Informatique de Paris~6, \textsc{LIP6},
    \'Equipe \textsc{PolSys} \\
\url{victor.magron@lip6.fr}
\and 
Mohab Safey El Din\\
Sorbonne Universit\'e, \textsc{CNRS}, \textsc{INRIA},\\
    Laboratoire d'Informatique de Paris~6, \textsc{LIP6},
    \'Equipe \textsc{PolSys} \\
\url{mohab.safey@lip6.fr}
}
\def\N{{\mathbb{N}}}
\def\Z{{\mathbb{Z}}}
\def\Q{{\mathbb{Q}}}
\def\R{{\mathbb{R}}}
\newcommand{\realcertify}{\textsc{RealCertify}}
\newcommand{\multivsos}{\texttt{multivsos}}
\newcommand{\univsos}{\texttt{univsos}}
\newcommand{\sumtwosquares}{\texttt{sum2squares}}
\newcommand{\univsosone}{\texttt{univsos1}}
\newcommand{\univsostwo}{\texttt{univsos2}}
\newcommand{\univsosthree}{\texttt{univsos3}}
\newcommand{\PP}{\texttt{RoundProject}}
\newcommand{\raglib}{\texttt{RAGLib}}
\newcommand{\cad}{\texttt{CAD}}
\newcommand{\sdp}{\texttt{sdp}}
\newcommand{\cholesky}{\texttt{cholesky}}

\date{}

\maketitle

\begin{abstract}
  Let $\Q$ (resp. $\R$) be the field of rational (resp. real) numbers and
  $X= (X_1, \ldots, X_n)$ be variables. Deciding the non-negativity of
  polynomials in $\Q[X]$ over $\R^n$ or over semi-algebraic
  domains defined by polynomial constraints in $\Q[X]$ is a
  classical algorithmic problem for symbolic computation. 

  The Maple package \textsc{RealCertify} tackles this decision problem by
  computing sum of squares certificates of non-negativity for inputs where such
  certificates hold over the rational numbers. It can be applied to numerous
  problems coming from engineering sciences, program verification and
  cyber-physical systems. It is based on hybrid symbolic-numeric algorithms
  based on semi-definite programming.
\end{abstract}

\section{Introduction}

Let $\Q$ (resp.~$\R$) be the field of rational (resp.~real) numbers and
$X = (X_1, \ldots, X_n)$ be a sequence of variables. We consider the problem of
deciding the non-negativity of $f \in \Q[X]$ either over $\R^n$ or over a
semi-algebraic set $S$ defined by some constraints
$g_1\geq 0, \ldots, g_m\geq 0$ (with $g_j \in \Q[X]$). We denote by $d$ the
maximum of the total degrees of these polynomials.

The Cylindrical Algebraic Decomposition (CAD) algorithm~\cite{Collins75} solves
this decision problem in time doubly exponential in $n$ (and polynomial in $d$).
This algorithm (and its further improvements) has been implemented in most of
computer algebra systems.

Later, the so-called critical point method has been designed, allowing to solve
this decision problem in time singly exponential in $n$ (and polynomial in $d$).
Recent variants of this method have also been implemented in the \href{http://www-polsys.lip6.fr/~safey/RAGLib/}{$\raglib$} Maple
package.

All the aforementioned algorithms are ``root finding'' ones: they try to find a
point at which $f$ is negative over the considered domain. When $f$ is positive,
they return an empty list without a {\it certificate} that can be checked {\it a
  posteriori}.

To compute certificates of non-negativity, an approach based on {\it sum of
  squares} (SOS) decompositions (and their variants) has been popularized by
Lasserre~\cite{Las01sos} and Parillo~\cite{phdParrilo}. The idea is as follows.

To ensure that a polynomial $f$ of degree $d = 2k$ is non-negative over $\R^n$, it suffices to
write it as a sum of squares $c_1 s_1^2+\cdots+c_r s_r^2$ where the $c_i$'s are
positive constants. When such $s_i$ and such $c_i$'s can be obtained with
rational coefficients, one says that one obtains a certificate of non-negativity
over the rationals. Such a decomposition can be obtained by finding a
semi-definite positive symmetric matrix $G$ such that
\[
f = v_k^T G v_k
\]
where $v_k$ is the vector of all monomials of degree $\leq k$. Obtaining such a
matrix $G$ boils down to solving a linear matrix inequality. 

This method is attractive because efficient numerical solvers are available for
solving {\em large} linear matrix inequalities. Besides, when $d$ is fixed and
$n$ grows, the size of the matrix $G$ varies polynomially in $n$, hence
providing {\em approximations} of a sum of squares decomposition for $f$. It
can also be generalized to obtain certificates of non-negativity for constrained
problems, writing $f$ as
\[
f = \sigma_0+ \sigma_1 g_1+\cdots + \sigma_m g_m
\]
where the $\sigma_i$'s are sum of squares.

On the minus side, this method provides only {\em approximations} of
certificates of non-negativity. Besides, it is well-known that not all
non-negative polynomials can be written as sum of squares of polynomials.
Original work of Parillo/Peyrl \cite{PaPe08} and Kaltofen/Li/Yang/Zhi
\cite{KLYZ08} have opened the door to hybrid symbolic numeric strategies for
computing certificates of non-negativity whenever such certificates exist over
the rational numbers.

In \cite{MaSa18}, we have designed hybrid symbolic-numeric algorithms for
computing certificates of non-negativity over the rationals in some ``easy''
situations (roughly speaking, these are the situations where the searched sum of
squares decomposition lies in the interior of the cone of polynomials which are
sum of squares). The package \textsc{RealCertify} implements these algorithms
and aims at providing a full suite of hybrid algorithms for computing
certificates of non-negativity based on numerical software for solving linear
matrix inequalities.
\section{Algorithmic background and overall description}

\subsection{The univariate case}
In the univariate case, all non-negative polynomials are sums-of-squares. The
library includes two distinct algorithms:
\begin{itemize}
\item $\univsosone$, which is a recursive procedure relying on root isolation
  and quadratic under approximations of positive polynomials.
  The first step computes a rational approximation $t$ of the smallest global
  minimizer $a$ of $f$ and a non-negative quadratic under-approximation $f_t$ of
  $f$ such that $t$ is a root of $f - f_t$. The second step performs square-free
  decomposition of $f - f_t = g h^2$. Then, we apply the same procedure on $g$
  until the resulting degree is less than 2.

\item $\univsostwo$, which relies on root isolation of perturbed positive
  polynomials. Given a univariate polynomial $f > 0$ of degree $d = 2k$, this
  algorithm computes weighted SOS decompositions of $f$. The first numeric step
  of $\univsostwo$ is to find $\varepsilon$ such that the perturbed polynomial
  $f_\varepsilon := f - \varepsilon \sum_{i=0}^k X^{2 i} > 0$ and to compute its
  complex roots, yielding an approximate SOS decomposition $l (s_1^2 + s_2^2)$,
  where $l$ is the leading coefficient of $f_\varepsilon$. In the second
  symbolic step, one considers the remainder polynomial
  $u := f_\varepsilon - l s_1^2 - l s_2^2$ and tries to computes an exact SOS
  decomposition of $\varepsilon \sum_{i=0}^k X^{2 i} + u$. This succeeds for
  large enough precision of the root isolation procedure.
%
%\item $\univsosthree$, which relies on sum of squares approximation (via semi-definite programming) of perturbed positive polynomials and square-free decomposition.
\end{itemize}
In both cases, the output is a list $[c_1,s_1,\dots,c_r,s_r]$, with
$c_i \in \Q^{> 0}$, $s_i \in \Q[X]$, such that
$f = c_1 s_1^2 + \dots + c_r s_r^2$.
Let us illustrate the behavior of both algorithms on the input
$f = 1 + X + X^2 + X^3 + X^4$.
\begin{enumerate}
\item When running~$\univsos1$, the algorithm first provides the value $t = -1$
  as an approximation of the minimizer of $f$ together with a positive quadratic
  under-approximation $f_{t}(X) = X^2$. Next, one obtains the square-free
  decomposition $f - f_t = (X+1)^2 g(X)$ with
  $g(X) = (X-\frac{1}{2})^2 + \frac{3}{4}$.
  The Maple command: \verb?univsos1(1+X+X^2+X^3+X^4,X)? outputs the list
  $[1, 0, 1, (X + 1) (X - \frac{1}{2}), \frac{3}{4}, X + 1, 1, -X]$,
  corresponding to the weighted rational SOS decomposition
  $f = (X+1)^2 (X-\frac{1}{2})^2 + \frac{3}{4} (X+1)^2 + (-X)^2$.
\item When running~$\univsos2$, the algorithm performs the first loop and
  provides the value $\varepsilon = \frac{1}{8}$ with the polynomial
  $f_\varepsilon := f - \frac{1}{8} (1 + X^2 + X^4)$ which has no real root. The
  leading coefficient of $f_\varepsilon$ is $l = \frac{7}{8}$.
  After multiplying the precision of complex root isolation by 8, one obtains
  $s_1 = X^2 + \frac{9}{16} X - \frac{3}{4}$,
  $s_2 = \frac{23}{16} X+ \frac{11}{16}$ and
  $u = \frac{1}{64} X^3 + \frac{105}{1024} X^2 + \frac{9}{1024} X -
  \frac{63}{2048}$.
  Using that $X^3 = \frac{1}{2} (X^2 + X)^2 - \frac{1}{2} (X^4 + X^2)$ and
  $X = \frac{1}{2} (X + 1)^2 - \frac{1}{2} (X^2 + 1)$, one gets an SOS
  decomposition for $u + \frac{1}{8} (1 + X^2 + X^4)$.
  The Maple command \verb?univsos2(1+X+X^2+X^3+X^4,X)? outputs the decomposition
  $f = \frac{7}{8}(s_1^2 + s_2^2) + \frac{377}{4096} + \frac{55}{256} X^2 +
  \frac{7}{64} X^4 + \frac{9}{1024} (X+\frac{1}{2})^2 + \frac{1}{64}
  X^2(X+\frac{1}{2})^2$.

\end{enumerate}

\subsection{The multivariate case}

In the multivariate case, the $\multivsos$ library performs SOS decompositions
of multivariate non-negative polynomials with rational coefficients in the
(un)-constrained case.

In the unconstrained case, $\multivsos$ implements a hybrid numeric-symbolic
algorithm computing exact rational SOS decompositions for polynomials lying in
the interior $\mathring{\Sigma}[X]$ of the SOS cone $\Sigma[X]$. It computes an
approximate SOS decomposition for a perturbation of the input polynomial with an
arbitrary-precision semi-definite programming (SDP) solver. An exact SOS
decomposition is obtained thanks to the perturbation terms.
Given $f \in \Z[X] \cap \mathring{\Sigma}[X]$ of degree $d = 2k$, one first
computes its Newton polytope $P$. The support of the SOS involved in the
decomposition of $f$ lies in $Q = P/2 \cap \N^n$.
A first loop allows to find $\varepsilon \in \Q^{>0}$ such that the perturbed
polynomial $f_\varepsilon := f - \varepsilon \sum_{\alpha \in Q} X^{2 \alpha}$
is also in $\Z[X] \cap \mathring{\Sigma}[X]$. In the second loop, one computes
an approximate rational SOS decomposition $\tilde{\sigma}$ of $f_\varepsilon$
with an arbitrary-precision SDP solver ($\sdp$ procedure). We obtain the
remainder
$u = f - \varepsilon \sum_{\alpha \in Q} X^{2 \alpha} - \tilde{\sigma}$. When
the precision is large enough, the last symbolic step allows to retrieve an
exact rational SOS decomposition of
$u + \varepsilon \sum_{\alpha \in Q} X^{2 \alpha}$.

In the constrained case, $\multivsos$ relies on a similar procedure to compute
weighted SOS decompositions for polynomials positive over basic compact
semi-algebraic sets.

We apply~$\multivsos$ on
$f = 4 X_1^4 + 4 X_1^3 X_2 - 7 X_1^2 X_2^2 - 2 X_1 X_2^3 + 10 X_2^4$. The other
input parameters are $\varepsilon = 1$, $\delta = R = 60$ and $\delta_c = 10$.
Then $Q := \conv{(\spt{f})}/2 \cap \N^n = \{(2,0),(1,1),(0,2)\}$. At the end of
the first loop, we get
$f - \varepsilon t = f - (X_1^4 + X_1^2 X_2^2 + X_2^2) \in
\mathring{\Sigma}[X]$.
The $\sdp$ and $\cholesky$ procedures yield
$s_1 = 2 X_1^2+ X_1 X_2- \frac{8}{3} X_2^2$,
$s_2 = \frac{4}{3} X_1 X_2+ \frac{3}{2} X_2^2$ and $s_3 = \frac{2}{7} X_2^2$.
The remainder polynomial is
$u = f- \varepsilon t - s_1^2 - s_2^2 - s_3^2 = - X_1^4- \frac{1}{9} X_1^2
X_2^2- \frac{2}{3} X_1 X_2^3- \frac{781}{1764} X_2^4$.

At the end of the second loop, we obtain
$\varepsilon_{(2,0)} = \varepsilon - X_1^4 = 0$, which is the coefficient of
$X_1^4$ in $\varepsilon t + u$. Then,
$\varepsilon (X_1^2 X_2^2 + X_2^4) - \frac{2}{3} X_1 X_2^3 = \frac{1}{3} (X_1
X_2 - X_2^2)^2 + (\varepsilon - \frac{1}{3}) (X_1^2 X_2^2 + X_2^4)$.
In the polynomial $\varepsilon t + u$, the coefficient of $X_1^2 X_2^2$ is
$\varepsilon_{(1,1)} = \varepsilon - \frac{1}{3} - \frac{1}{9} = \frac{5}{9}$
and the coefficient of $X_4^4$ is
$\varepsilon_{(0,2)} = \varepsilon - \frac{1}{3} - \frac{781}{1764} =
\frac{395}{1764}$.

The Maple command
\begin{verbatim}
multivsos(4 * X1^4 + 4 * X1^3 * X2 - 7 * X1^2 * X2^2 - 2 * X1 * X2^3 + 10 * X2^4):
\end{verbatim}

allows to obtain the weighted rational SOS decomposition: $4 X_1^4 + 4 X_1^3 X_2 - 7 X_1^2 X_2^2 - 2 X_1 X_2^3 + 10 X_2^4 = \frac{1}{3} (X_1 X_2-X_2^2)^2 + \frac{5}{9} (X_1 X_2)^2 + \frac{395}{1764} X_2^4 + (2 X_1^2+ X_1 X_2- \frac{8}{3} X_2^2)^2 + (\frac{4}{3} X_1 X_2+ \frac{3}{2}  X_2^2)^2+(\frac{2}{7} X_2^2)^2)$.

\subsection{Dependencies}
The $\realcertify$ software is available and maintained as a GitHub repository
at \href{https://gricad-gitlab.univ-grenoble-alpes.fr/magronv/RealCertify}{Gitlab}. The
$\univsos$ and $\multivsos$ libraries have been tested with Maple 2016.
$\univsos$ requires the external
\href{http://pari.math.u-bordeaux.fr/download.html}{PARIGP} software for
$\univsostwo$, as well as the external SDP
solvers~\href{https://sourceforge.net/projects/sdpa/files/sdpa/sdpa_7.3.8.tar.gz}{SDPA}
(double precision)
and~\href{https://sourceforge.net/projects/sdpa/files/sdpa-gmp/sdpa-gmp.7.1.3.src.20150320.tar.gz}{SDPA-GMP}~\cite{Nakata10GMP}
(arbitrary-precision).

In addition of SDPA and SDPA-GMP used for the $\sdp$ procedure, $\multivsos$
requires the Maple
package~\href{http://www.math.uwo.ca/faculty/franz/convex}{Convex}, by M. Franz,
to compute Newton polytopes.

\section{Performance analysis and limitations}

Timings, which we report on below, were obtained on an Intel Core i7-5600U CPU
(2.60 GHz) with 16Gb of RAM. Most of the time is spent in the $\sdp$ procedure
for all benchmarks. Those benchmarks are standard ones in the polynomial
optimization community. We report here only on multivariate problems. We refer
to \cite{univsos} for a performance analysis in the univariate case.

The table on the left below reports on unconstrained problems, while the one on
the right reports on constrained ones. It appears that on this class of
problems, \textsc{RealCertify} scales better than CAD-based software (the Maple
package implementing CAD) and \textsc{RAGlib}. It should be observed that these
examples can actually be decomposed into sums of squares quite easily.

%We use the Maple~\texttt{Convex}
%package\footnote{\url{http://www-home.math.uwo.ca/~mfranz/convex}} to compute
%Newton polytopes. 
%Our subroutine $\sdp$ relies on the arbitrary-precision solver
%SDPA-GMP~\cite{Nakata10GMP} and the $\cholesky$ procedure is implemented with
%the function~\texttt{LUDecomposition} available within Maple. 

% \begin{table}[!h]
% \begin{center}
% \caption{\footnotesize $\multivsos$ vs $\univsostwo$ \cite{univsos} for benchmarks from~\cite{Chevillard11}.}
% \begin{tabular}{lcr|rr|rr}
% \hline
% \multirow{2}{*}{Id} & \multirow{2}{*}{$d$} & \multirow{2}{*}{$\tau$ (bits)} & \multicolumn{2}{c|}{$\multivsos$} & \multicolumn{2}{c}{$\univsostwo$} \\
%  & & & $\tau_1$ (bits) & $t_1$ (s)  & $\tau_2$ (bits) & $t_2$ (s)  \\
% \hline  
% \# 1 & 13 & 22 682 & 387 178 & 0.84 & 51 992 & 0.83  \\
% \# 3 & 32 & 269 958 & $-$ & $-$ & 580 335 & 2.64  \\
% \# 4 & 22 & 47 019 &  1 229 036 & 2.08  & 106 797 & 1.78 \\
% \# 5 & 34 & 117 307 & 10 271 899 & 69.3  &   265 330 & 5.21 \\
% \# 6 & 17 & 26 438 & 713 865 & 1.15  & 59 926 & 1.03 \\
% \# 7 & 43 & 67 399 & 10 360 440 &  16.3  & 152 277 & 11.2  \\
% \# 8 & 22 & 27 581 & 1 123 152 & 1.95 & 63 630 & 1.86 \\
% \# 9 & 20 & 30 414 & 896 342 & 1.54  & 68 664 & 1.61 \\
% \# 10 & 25 & 42 749 & 2 436 703 & 3.02  &  98 926 & 2.76  \\
% \hline
% \end{tabular}
% %
% \label{table:bench1}
% \end{center}
% \end{table}
\vspace{-0.4cm}
{\tiny
  \begin{center}
\begin{tabular}{ll}
  \begin{minipage}{0.5\linewidth}
%\begin{table}[!ht]
\begin{center}
%\caption{\footnotesize $\multivsos$ vs $\raglib$ vs $\cad$ (Polya).
% for $n$-variate polynomials of degree $d$.
%}
%\vspace*{-0.4cm}
 \begin{tabular}{|lrr|rr|rr|c|c|}
\hline
\multirow{2}{*}{Id} & \multirow{2}{*}{$n$} & \multirow{2}{*}{$d$}  & \multicolumn{2}{c|}{$\multivsos$} &  $\raglib$ & $\cad$ \\
 & & & $\tau_1$ (bits) & $t_1$ (s)  & $t_3$ (s) & $t_4$ (s) \\
\hline  
$f_{12}$ & 2 & 12 & 162 861 & 5.96 & 0.15 & 0.07 \\
$f_{20}$ & 2 & 20 & 745 419 & 110. & 0.16 & 0.03 \\
$M_{20}$ & 3 & 8 & 4 695 & 0.18 &  0.13 & 0.05 \\
$M_{100}$ & 3 & 8 & 17 232 & 0.35 &  0.15 & 0.03 \\
$r_2$ & 2 & 4 & 1 866 & 0.03 &  0.09 & 0.01 \\
$r_4$ & 4 & 4 & 14 571 & 0.15  & 0.32 & $-$ \\
$r_6$ & 6 & 4 & 56 890 & 0.34  & 623. & $-$ \\
$r_8$ & 8 & 4 & 157 583 & 0.96  & $-$ & $-$ \\
$r_{10}$ & 10 & 4 & 344 347 & 2.45  & $-$ & $-$ \\
$r_6^2$ & 6 & 8 & 1 283 982 & 13.8  & 10.9 & $-$ \\
\hline
\end{tabular}
\label{table:bench2}
\end{center}
%\end{table}    
  \end{minipage}
  &
    \begin{minipage}{0.5\linewidth}
%\begin{table}[!ht]
\begin{center}
%\caption{\footnotesize $\multivsos$ vs $\raglib$ vs $\cad$ (Putinar).}
%\vspace*{-0.4cm}
 \begin{tabular}{|lrr|rrr|c|c|}
\hline
\multirow{2}{*}{Id} & \multirow{2}{*}{$n$} & \multirow{2}{*}{$d$}  & \multicolumn{3}{c|}{$\multivsos$} & $\raglib$ & $\cad$ \\
 & & & $k$ & $\tau_1$ (bits) & $t_1$ (s) & $t_2$ (s) & $t_3$ (s) \\
\hline  
$p_{46}$ & 2 & 4 & 3 & 21 723 & 0.83 & 0.15 & 0.81 \\
$f_{260}$ & 6 & 3 & 2 & 114 642 & 2.72 & 0.12 & $-$ \\
$f_{491}$ & 6 & 3 & 2 & 108 359& 9.65 & 0.01 & 0.05 \\
$f_{752}$ & 6 & 2 & 2 & 10 204 & 0.26 & 0.07 & $-$ \\
$f_{859}$ & 6 & 7 & 4 & 6 355 724 & 303. & 5896. & $-$ \\
$f_{863}$ & 4 & 2 & 1 & 5 492 & 0.14 & 0.01 & 0.01 \\
$f_{884}$ & 4 & 4 & 3 & 300 784 & 25.1 & 0.21 & $-$ \\
$f_{890}$ & 4 & 4 & 2 & 60 787 & 0.59 & 0.08 & $-$ \\
butcher & 6 & 3 & 2 & 247 623 & 1.32 & 47.2 & $-$ \\
heart & 8 & 4 & 2 &  618 847 & 2.94 & 0.54 & $-$ \\
magn. & 7 & 2 & 1 & 9 622 & 0.29 & 434. & $-$ \\
\hline
\end{tabular}
\label{table:bench3}
\end{center}
%\end{table}      
    \end{minipage}
\end{tabular}    
  \end{center}
}

The technique on which \textsc{RealCertify} relies takes also plenty advantage
on the fact that solving linear matrix inequalities at fixed precision can be
done in polynomial time when $d$ is fixed and $n$ increases.

However, we mention that for non-negative polynomials which are not sums of
squares, or which have coefficients with large magnitude, the practical
behaviour of \textsc{RealCertify} can be much less satisfactory and less
efficient than e.g. \textsc{RAGlib}.

Hence, one can see \textsc{RealCertify} as to be used in a pre-process for
getting cecrtificates of non-negativity which can be completed with other
symbolic computation tools.


\begin{thebibliography}{99}

\bibitem{Chevillard11}
S.~Chevillard, J.~Harrison, M.~Joldes, and C.~Lauter.
\newblock Efficient and accurate computation of upper bounds of approximation
  errors.
\newblock {\em Theoretical Computer Science}, 412(16):1523 -- 1543, 2011.

\bibitem{Collins75}
G.~E Collins.
\newblock Quantifier elimination for real closed fields by cylindrical
  algebraic decompostion.
\newblock In {\em ATFL 2nd GI Conf. Kaiserslautern}, pages 134--183, 1975.

\bibitem{KLYZ08}
E.~Kaltofen, B.~Li, Z.~Yang, and L.~Zhi.
\newblock Exact certification of global optimality of approximate
  factorizations via rationalizing sums-of-squares with floating point scalars.
\newblock In {\em Proceedings of the twenty-first international symposium on
  Symbolic and algebraic computation}, pages 155--164. ACM, 2008.

\bibitem{Las01sos}
J.-B. Lasserre.
\newblock {Global Optimization with Polynomials and the Problem of Moments}.
\newblock {\em SIAM Journal on Optimization}, 11(3):796--817, 2001.


\bibitem{MaSa18}
V.~Magron and M.~Safey El Din.
\newblock {On Exact Polya and Putinar's Representations}.
\newblock To appear in {\em Proceedings of the 2018 {ACM} International Symposium on Symbolic and Algebraic Computation (ISSAC)}.


\bibitem{univsos}
V.~Magron, M.~Safey~El Din, and M.~Schweighofer.
\newblock {Algorithms for Sums of Squares Decompositions of Non-negative
  Univariate Polynomials}, 2017.
\newblock Submitted.

\bibitem{Nakata10GMP}
M.~Nakata.
\newblock {A numerical evaluation of highly accurate multiple-precision
  arithmetic version of semidefinite programming solver: SDPA-GMP, -QD and
  -DD.}
\newblock In {\em CACSD}, pages 29--34, 2010.


\bibitem{phdParrilo}
{P. A. Parrilo}.
\newblock {\em {Structured Semidefinite Programs and Semialgebraic Geometry
  Methods in Robustness and Optimization}}.
\newblock PhD thesis, California Inst.~Tech., 2000.

\bibitem{PaPe08}
H.~Peyrl and P.A. Parrilo.
\newblock Computing sum of squares decompositions with rational coefficients.
\newblock {\em Theoretical Computer Science}, 409(2):269--281, 2008.


\end{thebibliography}
\end{document}